# Metabolic scaling law for fetus and placenta

Carolyn M. Salafia, Michael Yampolsky

June 1, 2008

**Kleiber's Law for scaling of basal metabolic rate.**

Kleiber's Law is named after the ground-breaking work of Max Kleiber (1),(2) in the early 1930's, who postulated that the basal metabolic rate $B$ of an organism is proportional to its mass $M$ raised to the power $3/4$:

$$B \sim M^{3/4}.$$

This law, discussed in detail in Kleiber's book *"The Fire of Life"* (1961) (3) has attracted both attention and controversy. The prevalent theory at that time (which has not yet been put completely to rest) was known as the *surface theory* of metabolism. It appeals to seemingly simple geometric considerations, that suggest the scaling exponent $2/3$ (rather than $3/4$). For example, let us make the simplifying assumptions that the body of an organism is a three-dimensional ball, and all of its metabolic exchange occurs via heat transfer through its spherical surface. Then the mass is proportional to the volume of the ball, given by $V = \frac{4}{3}\pi r^3$, and the basal metabolic rate is proportional to the surface area $A = 4\pi r^2$. Thus the scaling equation under these assumptions would be

$$B \sim M^{2/3}.$$

A similar argument has to do with the internal surfaces of the organism, where the oxygen and nutrient exchange takes place – they are again assumed to be proportional to the whole body surface.

However, Kleiber found that $3/4$ was generally in a much better agreement with experiment, as we confirm below. There are many errors in the arguments in favor of surface theory. A detailed refutation is found in Kleiber's book (3). To understand why metabolic rate is higher than the surface theory predicts, let us concentrate on the argument about the internal surfaces. To begin with, the circulatory system of an organism, where oxygen and nutrient exchange takes place, is not a ball, but rather a complicated network of vessels with a *fractal* structure. Metabolic rate in such a network would not have a scaling constant $2/3$.

To see how the fractal structure of the circulatory network potentially impacts the metabolic rate, let us look at a toy model of a two-dimensional organism, whose mass is proportional to the surface area, and whose basal metabolic rate measures energy exchange through its one-dimensional boundary (the "surface"), and is thus proportional to the perimeter.

As first example, consider an organism shaped as a circle. Given that the area scales as $r^2$, and the perimeter scales as $r$, we will have

$$B \sim M^{1/2}.$$

As a second example, consider an organism shaped as a simple fractal, known as *Koch snowflake*. To construct this shape, take an equilateral triangle of side $a$. Attach an equilateral triangle of side $a/3$ in the middle of each of the three sides (Figure 1, Step 1). Now repeat this procedure, attaching a triangle of side $a/9$ in the middle of every boundary segment, and so on (Figure 1, Steps 2 and 3). Note, that at every step of the construction, the perimeter increases by a multiple $4/3$, so if continued indefinitely, the process will yield a shape with an infinite perimeter. Of course, this would not be biologically realistic, and so we will stop adding new decorations to the snowflake at the step $n$, when the size of the new decoration $a/3^n$ becomes smaller than some cutoff value $l_0$.

**Figure 1.** A few steps in the construction of Koch Snowflake

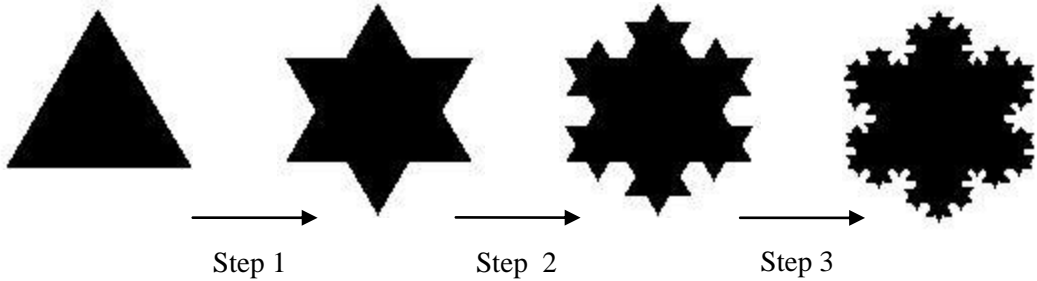

Step 1    Step 2    Step 3

By contrast, adding a new round of decorations has little effect on the *area*. An easy calculation shows that the total area of all new triangles added at step $n$ is

$$a^2 \frac{3\sqrt{3}}{16} \left(\frac{4}{9}\right)^{n-1}.$$

This quantity shrinks to zero rapidly as $n$ increases, and the total area converges to a finite number

$$L_a = \frac{2\sqrt{3}a^2}{5}.$$

To understand the scaling relation between the area $A_a$ and the perimeter $P_a$ of our toy organism, let us increase the length $a$ by a convenient multiple of 3. The new object will not only be bigger

– it will also have one more generation of decorations (bigger than $l_0$). As we know, the new area would be very near the value of $L_{3a}$, so

$$A_{3a} \approx 9 A_a.$$

But the increase in the perimeter will reflect the new decorations:

$$P_{3a} = \frac{4}{3} 3 P_a.$$

Assuming a scaling relation $P \sim A^\alpha$, we have

$$\log P_{3a} - \log P_a = \alpha(\log A_{3a} - \log A_a).$$

Solving for α, we get

$$\alpha = \frac{\log 4}{\log 9} = \log_9 4 \approx 0.63.$$

Compare this with our first two-dimensional example of a circular shape, for which α=0.5. Thus, the fractal structure of the perimeter leads to a *larger* scaling constant α in the relation

$$B \sim M^\alpha.$$

In 1997, G.B. West, J. H. Brown, and B. J. Enquist (4) proposed an explanation of Kleiber's scaling exponent based on the fractal nature of the vasculature. After making several simplifying assumptions, they derive $3/4$ from considerations of hydrodynamic optimality of the vascular network. This work has attracted a lot of criticism (see (5) as an example). However, the work of West *et al* appears to us as a step in the right direction, even if the model details are arguable. The explanation of a non-trivial scaling law based on spontaneously emerging fractal structure of an organism is consistent with well-understood examples in physics (see an excellent review (6) for other physical and biological examples). Indeed, one of the weaker points in (4) in our view, is that West *et al* assume a very *regular* fractal structure of the vasculature, when reality should be even more complex.

It is important to note, that at least some of the controversy around $3/4$ scaling can be attributed to the difficulty in *defining and measuring* the basal metabolic rate of an organism. Metabolism was historically thought of as heat exchange (as noted from the units in which metabolic rate is measured – *calories/second*). On the other hand, the analysis of West *et al*, for example, is entirely based on blood flow considerations.

**Kleiber's Law for human placenta.**

In 1966 Ahern, as cited in (7), proposed a variant of metabolic scaling law for human pregnancy. He proposed replacing the basal metabolic rate of a fetus with the placental mass *PM*, which should result in a scaling relation of the following kind (here *FM* stands for the body mass of the fetus):

$$PM \sim FM^{\beta}.$$

This step is hugely important, for two reasons. First, it removes the main obstacle in exactly defining what constitutes metabolic exchange, and measuring basal metabolic rate: *whatever energy is exchanged by the fetus must pass through the placenta.* Secondly, it leads to a measurement of the basal metabolic rate which does not require sophisticated laboratory apparatus to gauge the caloric intake and expenditure of an organism. Note that the measurements used in Kleiber's original work used tens (if not less) organisms of each kind. Ahern conjectured $\beta = 2/3$ for reasons already familiar to us.

Of course, *FM* is routinely recorded at birth. *PM* is also frequently recorded at birth, and is often combined with *FM* to yield the measurement known as *fetoplacental ratio (FPR)*. This is defined either as ratio of the mass of the placenta to that of the baby, or its inverse. It is commonly used for analysis of fetal pathologies, see for instance, the study in (8). To fix the ideas, we will set

$$FPR = FM/PM \text{ measured at birth.}$$

Recently, in (9), C.M. Salafia *et al* have studied the scaling relation between *PM* and *FM* using the data collected by the National Collaborative Perinatal Project. In all, over 26 thousand (!) pairs *(FM, PM)* were recorded. The results are in a remarkable agreement with Kleiber's Law:

$$PM = \alpha \cdot FM^{\beta}, \text{ with } \beta = 0.78 \pm 0.02, \alpha = 1.03 \pm 0.17$$

We should make a note, which Kleiber makes explicit in his book (3) on page 214, that until there is an accepted explanation of the specific value of the scaling exponent (such as that proposed in (4)), we cannot assume that the value $3/4$ is exactly correct. While it is a "nice number", so is $\pi/4$ – and we would not be able to discern a measurable difference between the two from the data. However, the value $3/4$ is preferable from practical considerations, as it is easier to calculate. Having said this, we turn to the practical aspects of applying the scaling law.

**Consequences for measuring *FPR*.**

Apart from very strong evidence in favor of Kleiber's Law, the main practical consequence concerns measuring *FPR*. The scaling relation

$$PM \approx FM^{0.75}$$

implies that *FM* grows *faster* than *PM* in a normal fetal development (see graph below). For instance, a normally developed fetus will have a larger *FPR* later in gestation. Since a low *FPR* is usually taken as a sign of poorer fetal growth and likely placental pathology, the findings based on *FPR* will be biased to favor newborns with a larger *FM*. For instance, a normally developed baby with a birth weight of *3000* grams will have mean *FPR* of almost *7.5%* lower than a baby with a birth weight of *4000* grams (see Figure 2). The difference for babies born weighing *2000* and *4000* grams is almost *20%*. To correct for this, the *FPR* should be replaced with a ratio reflecting the true scaling exponent:

$$FPR_{corrected} = {FM^{0.75}}/{PM}.$$

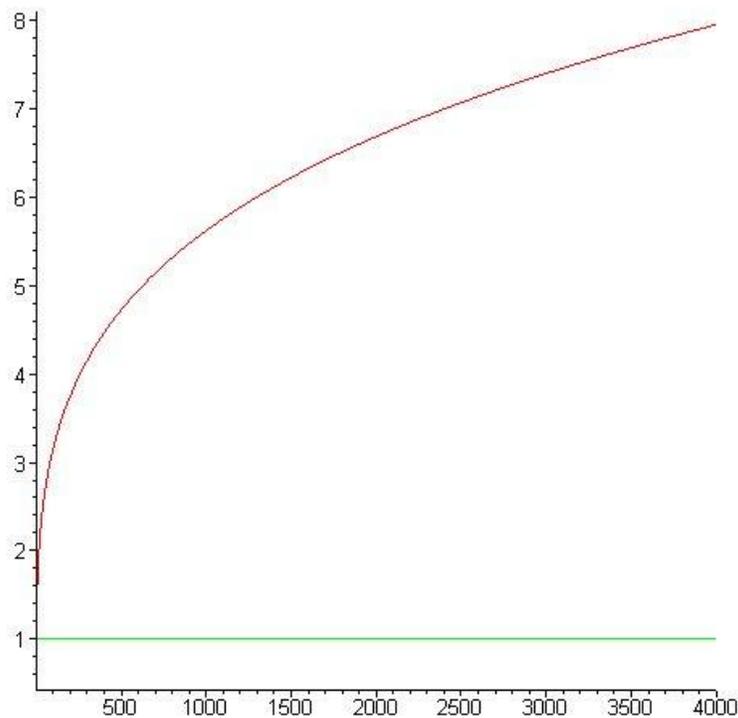

**Figure 2**. A sample graph of *FPR* (above) and *FPR$_{corrected}$* below for a normally developing fetus. The horizontal axis is *FM*. The bias in *FPR* is evident.

It is a coincidence that the mean value of $FPR_{corrected}$ obtained in our study is so close to *1*. There is no deep meaning to this, only a fortuitous choice of units for measuring masses. If we had elected to measure masses in ounces, for instance, the coefficient α would have the value of approximately *2.37*. In practice, however, it is useful, as it gives a simple prediction: *the normal value of $FPR_{corrected}$ is 1*.

**Correct scaling as an indicator of normal development.**

It is reasonable to assume, that a deviation from the ¾ scaling law would be associated with a variation in the normal fetal-placental development. A simple indicator of a normally developed placenta is its shape. A normal placenta is round, with a centrally inserted umbilical cord (10). In (11), we have studied correlation between violation of ¾ law, and non-roundness of the placenta. Two measures of placental shapes were used. The standard deviation of the radius of the placenta, calculated from the insertion point of the umbilical cord is the most obvious measure of roundness. The other one, which we call *roughness*, is calculated as the perimeter of the placental surface, divided by the perimeter of the smallest convex hull that contained the surface. It is equal to *1* for any convex shape, such as a circle, and thus measures deviation from convexity of the placental shape. The table of findings from (11) is shown below. Δβ is the difference from measured value of β and ¾.

**Table 1:** Correlation of the deviation from a round shape with a deviation from the ¾ rule. Lower value of significance level means stronger evidence (e.g. 0.009 means 9 in a 1000 chance that the correlation happens by a coincidence).

| | | |
|---|---|---|
| *Standard deviation of radius from insertion point* | Pearson Correlation with *Δβ* | **-.076** |
| | Significance | **.009** |
| *Roughness* | Pearson Correlation with *Δβ* | **.091** |
| | Significance | **.002** |
| *Standard deviation of radius from geometric center* | Pearson Correlation with *Δβ* | .020 |
| | Significance | .485 |

In (11) we speculate that a cause of the deformed placental shape is a variation in the development of the placental vascular tree. We develop a dynamical growth model of a placental vasculature, and demonstrate that a change in the *branching density* of the tree results in one of several typical deformed shapes. Looking back at the example of the Koch Snowflake, we see how a change in branching density would account for a change in the metabolic scaling law.

**Figure 3:** Typical deformed placental shapes from (11). Top row – placentas, bottom row – models of their vascular trees.

| Round shape | Star-like shape | Tri-lobate shape |
|---|---|---|
| 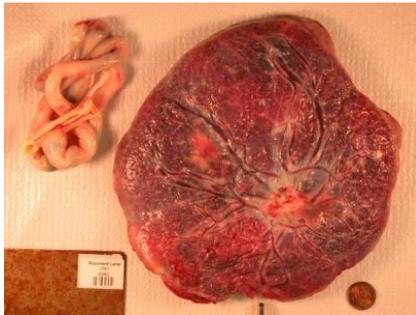 | 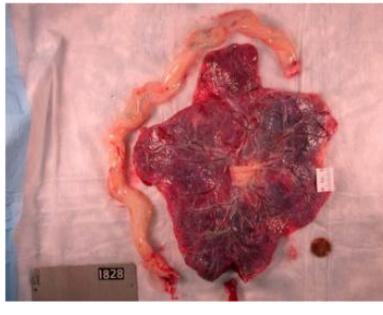 | 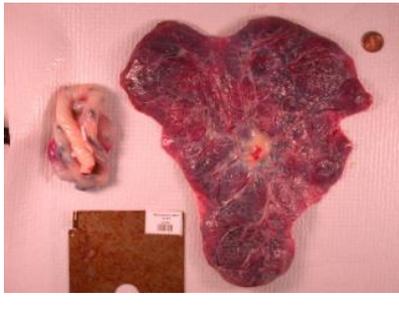 |
| 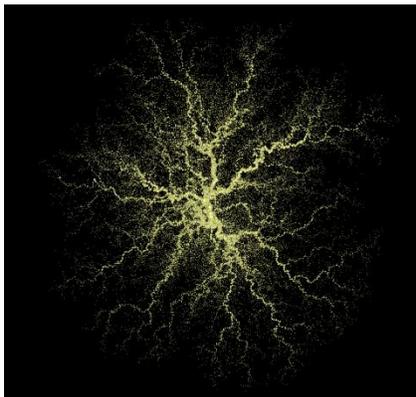 | 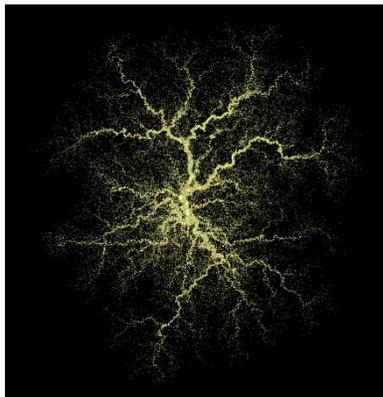 | 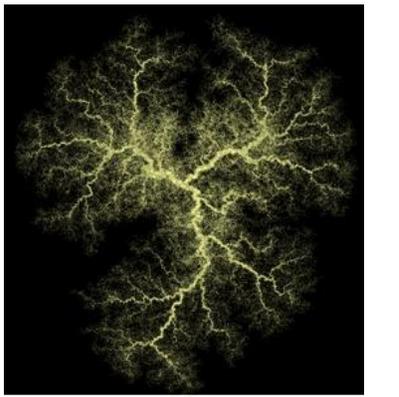 |

As a final piece of evidence here, consider that the standard deviation of radius from the *geometric center* of the placental surface (last row of the table) is not significantly correlated with Δβ. The geometric center has no particular significance in the internal structure of the placenta, whereas the insertion point of the umbilical cord is the root of the vascular tree. Thus we see that variation in the scaling is related to the deformations in the shape of the vascular tree, and not a deformed surface shape *per se*.

**The origins of metabolic scaling laws.**

There is a lively discussion in the existing literature of both the validity and origins of Kleiber's Law (as examples, see (4), (5)). Disagreements about the former could be partially attributed to the difficulty in measuring the basal metabolic rate, and the small sizes of the samples. In this regard, our study seems to conclusively confirm Kleiber's Law in fetal-placental setting. As to the origins of the scaling law, our model relates it to the structure of the vascular network. This is a partial confirmation of the results of (4). However, it should not be taken as conclusive validation of self-similar fractal models, such as that of (4), as our model exhibits a very complex scaling, more typical of self-similarity observed in physical systems. Rather, the model suggests that the scaling law has dynamical origins. Spontaneously occurring self-similarity (known as self-organized criticality (6)) in a complex dynamical system is a well-studied phenomenon in physics, and could be the direction to look for the origins of Kleiber's Law.

**Authors' affiliations:**

**C.S.**: Department of Psychiatry, New York University School of Medicine. 550 First Avenue, New York, NY 10016 and Department of Obstetrics and Gynecology, St Luke's Roosevelt Hospital, NY, NY, 10019.

**M.Y.**: Department of Mathematics, University of Toronto, 40 St. George St, Toronto, Ontario, Canada, M5S2E4

**Bibliography.**

1. *Body size and metabolism.* **Kleiber, M.** 1932, Hilgardia, Vol. 6, pp. 315-353.

2. *Body size and metabolic rate.* **Kleiber, M.** 1947, Physiol. rev., Vol. 27, pp. 511-541.

3. **Kleiber, M.** *The Fire of Life. Revised Ed.* s.l. : Robert E. Krieger Publ. Co., 1975.

4. *A general model for the origin of allometric scaling laws in biology.* **West, G.B., Brown, J.H. and Enquist, B. J.** 1997, Science, Vol. 276, pp. 122-126.

5. *Is West, Brown, and Enquist's model of allometric scaling methematically correct and biologically relevant?* **Kozlowski, J. and Konarzewski, M.** 2004, Functional Ecology, Vol. 18, pp. 283-289.

6. *Scale invariance in biology: coincidence or footprint of a universal mechanism.* **Gisiger, T.** 2001, Biol. Rev., Vol. 76, pp. 161-209.

7. **Gruenwald, P.** *The placenta and its maternal supply line: Effects of insufficiency on the fetus.* Baltimore : University Park Press, 1975.

8. *Gross morphological changes of placentas associated with intrauterine growth restriction of fetuses: A case control study.* **Biswas, S. and Ghosh, S.K.** 2007, Early Human Development .

9. *Allosteric metabolic scaling and fetal and placental weight.* **Salafia, C. M., et al.** 2007. Submitted for publication. .

10. **Benirschke, K.P.** Architecture of Normal Villous Trees. *Pathology of the Human Placenta.* New York : Springer Verlag, 2002, pp. 116-154.

11. *Modeling the variability of shapes of human placenta.* **Yampolsky, M., et al.** s.l. : Submitted., 2007.